\documentclass[%
 reprint,
 amsmath,
 amssymb,
 aps,
 twocolumn,
 superscriptaddress
]{revtex4-1}

\usepackage{upgreek}
\usepackage{verbatim}
\usepackage[normalem]{ulem}

\newcommand {\bise}{Bi$_2$Se$_3$}

\newcommand {\kf}{$k_\mathrm{F}$}
\newcommand {\ef}{$E_\mathrm{F}$}

\usepackage[colorlinks=True,linkcolor=red,citecolor=blue,urlcolor=blue]{hyperref}
\usepackage{amsmath}
\usepackage{graphicx}
\usepackage{bm}
\usepackage[caption=false]{subfig}
\usepackage{float}
\usepackage{placeins}
\usepackage{xcolor}
\usepackage{siunitx}
\usepackage{titlesec}
\usepackage{gensymb}
\usepackage{xcolor}
\usepackage{braket}

\titleformat{\section}
    {\normalfont\itshape\filcenter}{\thesection}{1em}{}
\titlespacing*{\section}{0pt}{0.5\baselineskip}{\baselineskip}

\begin{document}

\preprint{APS/123-QED}

\title{Establishing Coherent Momentum-Space Electronic States in Locally Ordered Materials}

\author
{Samuel T. Ciocys,$^{1,2*}$ Quentin Marsal,$^{5*}$ Paul Corbae,$^{2,3*}$ Daniel Varjas,$^{6}$ Ellis Kennedy,$^{3,4}$ Mary Scott,$^{3,4}$ Frances Hellman,$^{1,2}$ Adolfo G. Grushin,$^{5}$ and Alessandra Lanzara$^{1,2\ast}$\\
\normalsize{$^{1}$Department of Physics, University of California,}\\
\normalsize{Berkeley, California, 94720, USA}\\
\normalsize{$^{2}$Materials Science Division, Lawrence Berkeley National Laboratory,}\\
\normalsize{Berkeley, California, 94720, USA}\\
\normalsize{$^{3}$Department of Materials Science, University of California,}\\
\normalsize{Berkeley, California, 94720, USA}\\
\normalsize{$^{4}$
 Molecular Foundry, Lawrence Berkeley National Laboratory,}\\
\normalsize{Berkeley, California, 94720, USA}\\
\normalsize{$^{5}$Univ. Grenoble Alpes, CNRS, Grenoble INP, Institut N\'eel,}\\
\normalsize{38000 Grenoble, France}\\
\normalsize{$^{6}$Department of Physics, Stockholm University, AlbaNova University Center,}\\
\normalsize{106 91 Stockholm, Sweden}\\
\normalsize{$^\ast$To whom correspondence should be addressed; E-mail: alanzara@lbl.gov}
}

\date{\today}

\begin{abstract}
In our understanding of solids, the formation of highly spatially coherent electronic states, fundamental to command the quantum behavior of materials, relies on the existence of discrete translational symmetry of the crystalline lattice, a notion that has laid the foundation of Bloch’s theorem.  In contrast, in the absence of long-range order, as in the case of non-crystalline materials,
the electronic states are localized and electronic coherence does not develop. However, most current and future quantum technologies rely on materials that fall in between these two limits, where translational symmetry is broken, but short-range order with well-defined structural length scales persists. This brings forward the fundamental question whether long range order is necessary condition to establish coherence and structured momentum-dependent electronic state, and how to characterize it in the presence of short-range order. 
Here we study Bi$_2$Se$_3$, a material that exists in its crystalline form with long range order, in amorphous form, with short and medium range order, and in its nanocrystalline form, with reduced short range order.
By using angle resolved photoemission spectroscopy to directly access the electronic states in a momentum resolved manner, we reveal that, even in the absence of long-range order, a well-defined real-space length scale is sufficient to produce dispersive band structures. Moreover, we observe for the first time a repeated Fermi surface structure of duplicated annuli, reminiscent of Brillouin zone-like repetitions. These results, together with our simulations using amorphous Hamiltonians, reveal that the typical momentum scale where coherence occurs is the inverse average nearest-neighbor distance, the direct fingerprint of the local order of the underlying atomic structure. These results, not only lead the way to a new understanding of electronic coherence in solids, but also open the way to the realization of novel momentum- dependent quantum phenomena such as momentum pairing and spin-orbit coupling, in a much broader class of materials than the currently studied ones, lacking long range crystalline translational symmetry.

\end{abstract}

\maketitle

\section*{Introduction}

Electronic coherence is of utmost importance for the access and control of quantum-mechanical properties of materials. The foundation of our understanding of materials and modern technology is the realization that long-range order, in the form of discrete translational symmetry of the crystal lattice, is fundamental in establishing highly spatially coherent electronic states. In a crystalline solid, translational symmetry leads to well-defined peaks in the lattice structure factor, Fig.~\ref{fig:fig1}(a) top, that determine where Bragg scattering of electrons occurs. The scattering planes are determined by the periodic atomic potential, and define the edges of the Brillouin zone. Through Bloch’s theorem, the crystalline momentum precisely describes energy- and momentum-dependent electronic states confined to periodic Brillouin zones in momentum space. These electronic states are what ultimately define the materials ground state and the its transport, optical, magnetic and topological properties \cite{Ashcroft76}.

In contrast, in the case of non-crystalline solids, the Brillouin zone description is less clear due to the absence of long-range order. For instance, the structure factor of random atomic positions 
is uniformly distributed (Fig.~\ref{fig:fig1}(a) middle) and lattice disorder is expected to localize the electronic states, leading to a structureless dispersion relation (i.e. featureless momentum-space) \cite{Anderson1958,Mott1978,Ludlam2005, Mitchell2018}.

Much of our world however, is comprised of materials that fall in between these two categories, such as amorphous materials, high entropy alloys, quasicrystals, and liquid metals \cite{Varjas2019,Robarts2020,Ding2015}. They lack periodicity and long-range order, but still retain short-range ordering (SRO) with well-defined structural length scales, such as bond-lengths and preferred local environments. Structurally, this situation results in an atomic arrangement which is locally similar to the crystalline case (bond lengths, angles, and coordination), but globally the atomic sites demonstrate no periodic behavior \cite{zallen,Zachariasen1932,Vink2001,Deringer2018}. In this case, the diffraction pattern is not uniformly distributed, but rather presents a set of rings (Fig.~\ref{fig:fig1}(a) bottom) each corresponding to characteristic real-space scales, such as a well-defined nearest neighbour distance. These materials lie at the core of most modern technologies, bringing forward the fundamental questions of whether SRO by itself is a sufficient condition for establishing coherent and structured momentum-dependent electronic states, and how to characterize the resulting electronic structure. Although to date no direct study exists, the recent experimental observation of strongly dispersive surface states in a purely amorphous \bise{} (a-\bise{})\cite{Corbae2023}, and of a pseudogap-like band structure in the presence of disordered dopants on the surface of crystalline black phosphorus \cite{Ryu2021}, challenge the necessity of a Bloch theorem foundation for coherent electronic structure in momentum-space.

In this work, we use angle resolved photoemission spectroscopy (ARPES) to directly map the evolution of the electronic structure of \bise{} in momentum space, when it departs from its crystalline form. Specifically, we probe \bise{} in three unique structural forms. The first, crystalline \bise{}, has long-range order and translation symmetry. Amorphous \bise{} has well-defined short-range ordered bond-lengths and preferred local environments, even medium range ordering (MRO). Finally, nanocrystalline \bise{} has reduced SRO compared to amorphous systems since grain boundaries lack any SRO and no MRO.
The momentum resolving power of ARPES combined with the ability to control local-ordering during material synthesis place us in a prime position to address these questions directly.

Our results reveal the presence of a dispersive electronic structure and distinct Fermi surface in a-\bise{} with Brillouin-like zone repetitions. In contrast, in the case of nanocrystalline \bise{}, highly non-dispersive features are observed, likely due to scattering by the highly disordered atomic structure found at nanocrystal grain boundaries. The stark difference between the two, highlights the importance of the local environment for establishing coherence and structured momentum-dependent electronic states. These results coupled with simulations using amorphous Hamiltonians reveal that the typical momentum scale where repetitions occur is the inverse average nearest-neighbor distance, demonstrating that, even in the absence of long-range order, a well-defined real-space length scale is sufficient to produce dispersive band structures. This phenomenology is strongly reminiscent of the decades-old predictions concerning the dispersion of electrons in liquid metals \cite{Ziman1961,Edwards1961,Edwards1962,Morgan1969,Gyorffy1970,Schwartz1971,Chang1975,Olson1975,Oglesby1976}, where dispersive electronic features were predicted to remain contingent on the presence of a well-defined nearest neighbor distance. It also showcases that amorphous materials are in fact locally ordered systems, comparable to pristine crystalline systems, deserving further investigation in momentum space. Finally, given that many electronic phenomena rely on momentum- dependence, such as momentum pairing and spin-orbit coupling, our results put forward amorphous materials as a source for novel quantum phases of matter.

\section*{Results}

\begin{figure*}[ht]
    \includegraphics[width=1\textwidth]{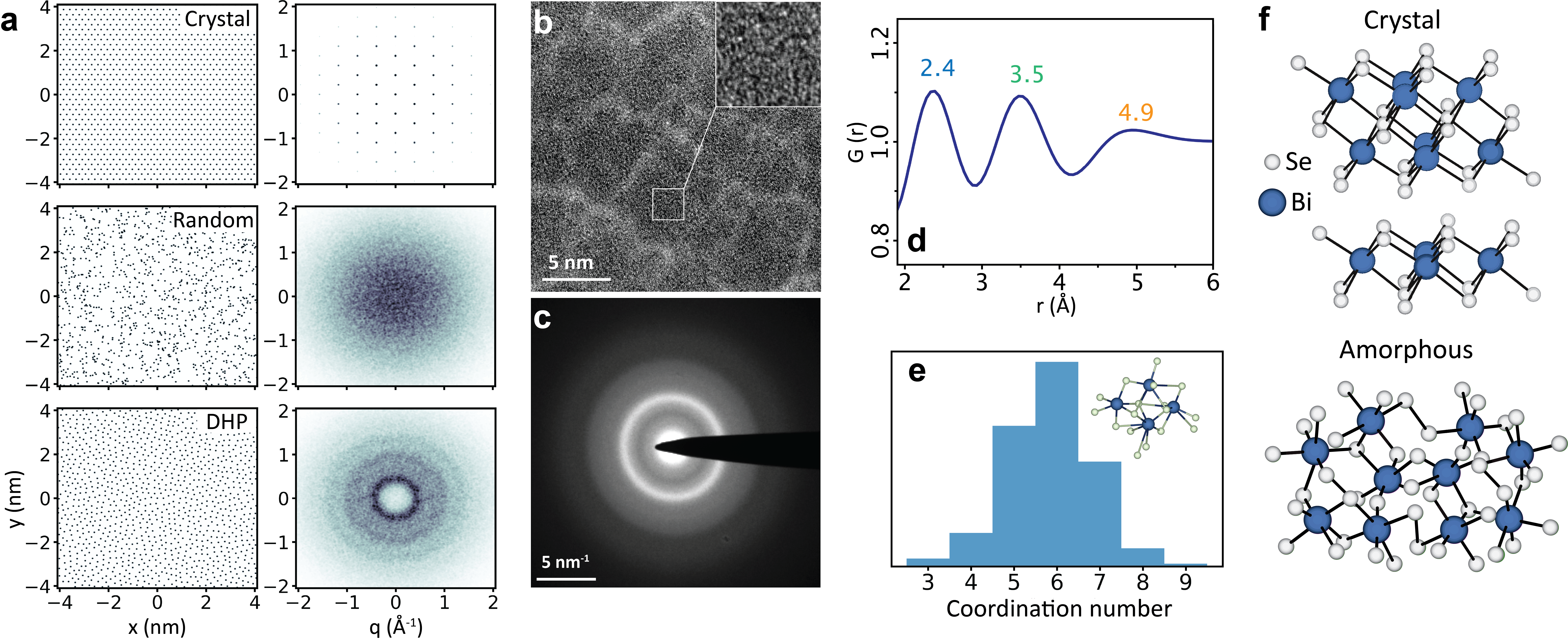}
    \caption{\textbf{Well-defined reciprocal length scale from real-space short-range order} \textbf{(a)} Fourier transforms for three real-space point distributions (crystalline, normal random, and disordered hard pack) demonstrates that reciprocal-space structure persists in the presence of well defined nearest-neighbor distance. \textbf{(b,c)} Large scale HRTEM image shows no signs of crystalline order like precursor lattice fringes. The electron diffraction pattern shows broad diffuse rings corresponding to SRO and no high intensity spots from long range order. \textbf{(d)} The reduced radial distribution function, $G(r)$, has three peaks from a well defined nearest neighbor ($2.4$ Å), next nearest neighbor ($3.5$ Å), and third nearest neighbor ($4.9$ Å). \textbf{(e)} Coordination number for amorphous \bise{} calculated using a 200 atom cell and ab-intio molecular dynamics. The CN is peaked at 6. Inset: an example coordination environment in amorphous \bise{}. \textbf{(f)} Ball-and-stick model of crystalline and amorphous \bise{}. For the amorphous structure, van der Waals separation is absent and majority sites are octahedral coordinated, implying an isotropic nearest neighbor distance.}
    \label{fig:fig1}
\end{figure*}

To begin, it is useful to recall that crystalline \bise{} (c-\bise{}) features a quintuple layer structure of alternating selenium and bismuth planes with bismuth atoms octohedrally coordinated with six adjacent selenium atoms. The quintuple layers are bonded by van der Waals forces, the stacking of which defines the c-axis lattice constant (see top panel in Fig.~\ref{fig:fig1} (f)).

The accumulated knowledge on amorphous systems suggests that a realizable structure for a-\bise{} can share certain traits with c-\bise{}. For example, in elemental amorphous materials, such as Si, Ge and monolayer carbon, or bi-elemental amorphous compounds such as SiO$_2$ and GaAs, the coordination of atoms and the nearest neighbor distances remain peaked at the values of their crystalline counterparts. The structural disorder stems from small variations in bond angles and smaller variations in bond lengths, which are peaked at the crystalline values\cite{Deringer2018,Ridgway1998,Onodera2020,toh2020}. Following suit, a-\bise{} is expected to also possess octohedrally coordinated bismuth atoms and similar local environment to c-\bise{}. The propensity for amorphous systems to retain the crystalline local order means that the amorphous system has a tendency to retain a well defined length-scale. A notable difference between the c-\bise{} and a-\bise{} however is that the van der Waals gap in c-\bise{} is an inherently 2D structure with no obvious analog in the amorphous case. Indeed, in Ref.~\cite{Corbae2023} we have demonstrated through Raman spectroscopy that the van der Waals gap no longer exists in a-\bise{}. Using ab-initio molecular dynamics to generate realistic a-\bise{} amorphous structures, we observe a peak in the coordination number at six (Fig. \ref{fig:fig1}(e)), representing the existence of majority octahedral environments.

To elucidate the real-space structure, we grew a-\bise{} using physical vapor deposition from two elemental effusion cells and characterized the structure using high-resolution transmission electron microscopy (HRTEM) in shown in Fig. \ref{fig:fig1}(b). The large scale HRTEM image indicates no regions exhibiting crystalline order or even nano-crystalline precursors (the contrast visible in the main image is associated with columnar microstructure that is common in thermally evaporated amorphous materials). The inset displays an expanded 2 nm x 2 nm field of view displaying a pattern due to phase contrast resulting from the lack of long-range periodicity, but has no sign of any nanocrystalline or even precursor nanocrystallites.
 
Panel (c) in Fig.~\ref{fig:fig1} shows TEM diffraction from the same film exhibiting the characteristic diffuse rings of amorphous systems lacking long range order. The presence of rings is indicative of well-preserved real-space length-scales. Using parallel-beam diffraction, we compute the reduced radial distribution function, $G(r)$, shown in Fig.~\ref{fig:fig1}(d). The a-\bise{} film shows clear peaks at $2.4$ Å, $3.5$ Å, and $4.9$ Å, indicating well-defined real and reciprocal length-scales in the system. 

\begin{figure*}[ht]
    \includegraphics[width=1\textwidth]{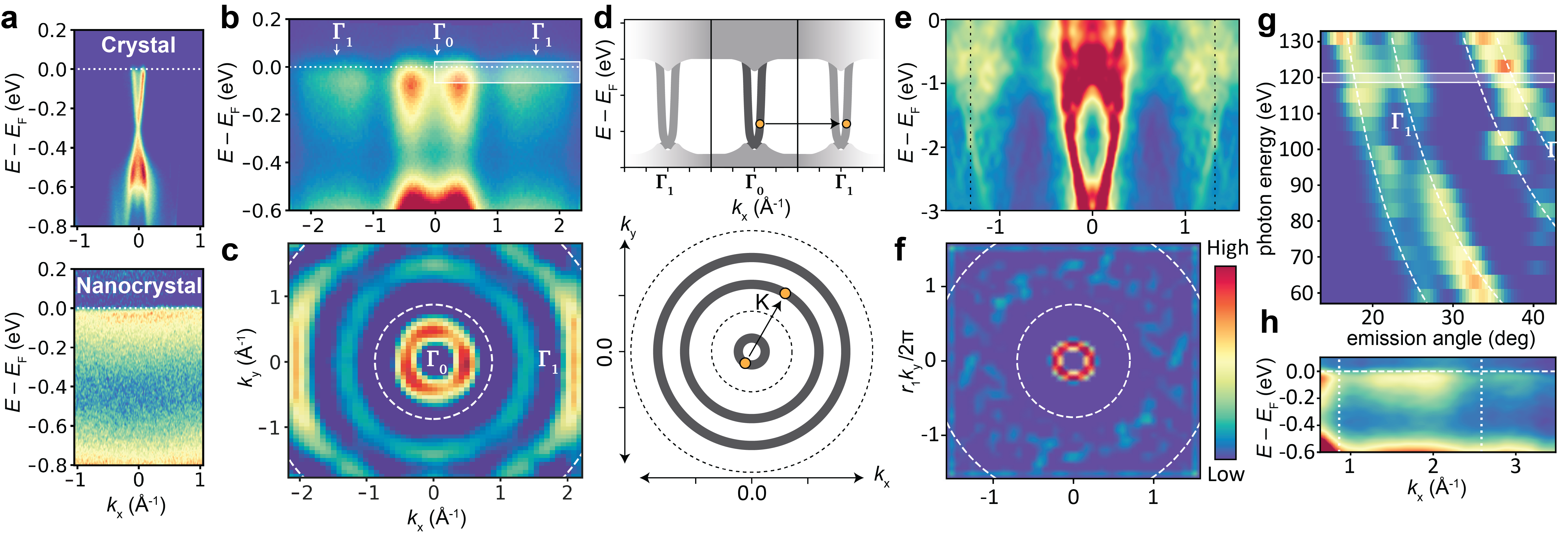}
    \caption{\textbf{Fermiology of the amorphous surface state.} \textbf{(a)} \textcolor{black}{ARPES spectra of crystalline and nanocrystalline \bise.} \textbf{(b)} Large momentum range ARPES spectrum of amorphous \bise uncovers duplicate dispersions approximately 1.75 Å$^{-1}$ from $\Upgamma$. \textbf{(c)} The Fermi surface ($\nabla^2I$ of the raw intensity for visibility) demonstrates rotational symmetry of the primary and duplicated dispersion. \textbf{(d)} Illustration of amorphous dispersion and Brillouin zone-like repetition contingent on a characteristic momentum. \textbf{(e)} Simulated dispersion along $k_\parallel$ through $\Upgamma$ showing duplicated structures. \textbf{(f)} Fermi surface from simulations showing repeated annuli. \textbf{(g)} Photon energy dependence of the 2nd and 3rd BZ dispersion, obeying $k_z$-independent photoemission (dashed white). \textbf{(h)} ARPES spectrum at $h\nu=120$ eV with repeated dispersions separated by 1.75 Å$^{-1}$.}
    \label{fig:fig2}
\end{figure*}

Figure \ref{fig:fig2} summarizes the momentum-space structure from the electronic dispersion we obtain from ARPES on a-\bise{}. 
For comparison with the amorphous spectrum in panels (b)-(h), panel (a) displays ARPES spectra for c-\bise{} and nanocrystalline \bise{} at photon energies of 115 eV and 100 eV, respectively. The crystalline sample exhibits a Dirac surface state and increased spectral intensity at the valence band near $E-E_{\mathrm{F}} = -0.6$ eV. The nanocrystalline sample is momentum-independent, with the only energy-dependent feature above $-1.0$ eV being reduced spectral intensity at $E-E_{\mathrm{F}} = -0.4$ eV.

Panel (b) displays the a-\bise{} spectrum along a momentum-slice cutting through $(k_x,k_y) = (0,0)$ at a photon energy of 120 eV, revealing remarkably dispersive band structure manifesting as vertical column-like features and an M-shaped valence band~\cite{Corbae2023}. The near-vertical features are in stark contrast to the expectation that disorder and localization lead to a broadened, momentum-independent electronic dispersion \cite{Anderson1958,Mott1978,Ludlam2005}, as seen in the nanocrystalline case. The clear difference between the amorphous and nanocrystalline spectra highlights how important the local environment is for electronic properties, with the possibility that grain boundaries are playing a substantial role in electronic decoherence.

Looking further at larger momenta, the band structure is replicated, resulting in copies of the electronic states at $-1.75$ and $1.75$ Å$^{-1}$ with reduced intensity. The replicas occur at a characteristic momentum, $k^{*} = 2\pi/a^{*}\approx 1.75$ Å$^{-1}$ corresponding to $a^{*}\approx 3.6$ Å, which closely matches the second peak in the radial distribution function from Fig.~\ref{fig:fig1}(d). The consequent Fermi surface, seen in Fig.~\ref{fig:fig2}(c) and visually enhanced by taking $\nabla^2I$ (where $I$ is photoemission intensity), confirms that the momentum space structure is rotationally symmetric in the form of concentric rings. Therefore, the dispersive structure is only repeated along the radial direction, forming annular regions at larger momenta.

The repetition phenomenon is reminiscent of that occurring in crystalline systems, which feature duplicated dispersions commensurate with reciprocal lattice vectors, outside of the first Brillouin zone. Hence, we refer to the regions where duplicates appear as "Brillouin-like zones" (BLZ) because they demonstrate repetition akin to crystalline fermiology. Unlike for crystals, the repetitions occur only along the radial direction, which we interpret as a manifestation of the rotational symmetry expected in amorphous structures. Fig.~\ref{fig:fig2}(d) conceptualizes the BLZ, showing the typical BZ-relationship of the replicated bands with respect to a reciprocal lattice constant (upper panel) and the annular zones in the Fermi surface (lower panel). The essential difference in the amorphous case is that its uniformity at long length-scales implies that reciprocal-space structure is rotational symmetric. Therefore reciprocal lattice vectors cannot exist, otherwise there would be well defined preferential directions and therefore long-range order. However preferential momentum scalars, can exist since local ordering, such as typical bond lengths, can persist in randomized directions. As a note, an effect which is absent in our spectra is the possible long-range nematic ordering. This can occur for instance via a compression or a strain which modifies the nearest-neighbor distances along one axis resulting in an elliptic transformation of the BLZ \cite{Ryu2021,Perret2021}. 

Expanding further out in momentum space exposes higher order BLZ. Fig.~\ref{fig:fig2}(g) shows the photon energy dependence of the dispersion at the Fermi level along the same radial direction as panel (b) at large detection angles. The white dashed lines represent the photon energy dependence for 2D states at four different $k_z$-independent momenta, indicating that the features are in fact photoemission from 2D surface states as opposed artifacts from photon energy-dependent matrix elements. The first two curves from the top-left follow the 1st order BLZ (0th order being at $\Upgamma_0$) and the last two curves follow the 2nd order BLZ. Panel (h) displays the momentum space converted spectrum for $h\nu = 120$ eV (shaded region in (g)) in which the bright 0th order dispersion is cutoff on the left edge and the 2nd order BLZ can be seen near 3 Å$^{-1}$. The intensity of the BLZs decrease and the dispersions broaden at larger momenta.

To determine the origin of the BLZ we compare the results of the ARPES experiment on a-Bi$_2$Se$_3$ to a numerical simulation with a tight-binding Hamiltonian of a-Bi$_2$Se$_3$ introduced in \cite{Corbae2023}. However, to explain our ARPES observations we need to ensure that a degree of local order is preserved when defining the atomic arrangement. To do so we construct a 3D arrangement of amorphous sites using thermalized hard packed spheres as in Ref.~\cite{Corbae2023} to which we add a relaxation step, resulting in a radial distribution function with peaks corresponding to nearest, second-nearest, and third-nearest neighbours. The peak locations define the characteristic length scales of the system. We use the nearest-neighbour average distance $r^{*}$ to set the scale of the plots in position space. This choice defines the characteristic momentum scale $k^{*} = \frac{2\pi}{r^{*}}$, which we use to normalize momentum space.

On this atomic site distribution, we define a model with a spin $\frac{1}{2}$ degree of freedom and two orbitals per site, to generalize the crystalline Bernevig-Hughes-Zhang (BHZ) model \cite{Zhang2009}, For each site in the amorphous structure we find the six closest neighbours. We then choose the coupling between neighbouring sites so that, if the lattice was cubic, it would result in the original BHZ model. Because the hoppings are assigned sequentially for every site, the six-fold coordination is preserved on average. The hopping term of the Hamiltonian depends on the relative position of the two sites $\mathbf{d}_{ij}$ according to
\begin{equation}
\label{eq:BHZ1}
    \braket{i|H|j} = it_1(\mathbf{\hat{d}}_{ij}\cdot\mathbf{\sigma})\tau_x-t_2\sigma_0\tau_z,
\end{equation}
while the onsite energy of a single site reads
\begin{equation}
    \braket{i|H|i} = M\left(\sigma_0\tau_z+\alpha e^{-\frac{\delta_i}{2r^{*}}}\sigma_0\tau_0\right). \label{eq:onsite}
\end{equation}
This system shows a topologically non-trivial gap that hosts a Dirac cone surface state for $M$ positive and close enough to zero\cite{Corbae2023}.  The parameter $\alpha$ controls the strength of a symmetry-allowed surface on-site potential that shifts the surface Dirac cone away from $E=0$, where $\delta_i$ is the distance from site $i$ to the surface.

We compute the spectral function of this system by projecting into a basis of plane waves of light momentum $k$, illuminating a single surface with a finite penetration depth, simulating an ARPES experiment. One can then define the two parallel and the perpendicular components of the momentum without ambiguity. Due to the isotropy of amorphous systems, the problem is invariant up to an in-plane rotation. One can thus only focus on the incident plane of light. The phenomenology we discuss next also applies to two-dimensional systems.

The ARPES spectra of a-\bise{} and the numerical model (Fig.~\ref{fig:fig2}(e)) both show the bulk gap, and within it a dispersive surface states that cross the gap around the $\Gamma$ point. Around each momenta commensurate with the characteristic momenta $k^{*}$, $\left \Vert \mathbf{k}_\parallel \right \Vert= k^{*}$, $2k^{*}$... a copy of the central bulk states and surface Dirac cone appears. This is the amorphous equivalent of the Brillouin zones of a crystal, enabled by the characteristic nearest-neighbour distance retained by the amorphous structure. This suggests that in Figs.~\ref{fig:fig2}(b) and (c), the repetitions we observe originate in the local order of the atomic sites. Fig.~\ref{fig:fig2}(f) shows that the expected spectrum is isotropic in the two components of the momentum that are parallel to the illuminated surface, as also observed experimentally. This theoretical analysis combined with the experimental spectra strongly supports the conclusion that ARPES can be used as a tool to extract the scale of local order of any non-crystalline solid by observing BLZ repetitions.

\begin{figure*}[ht]
    \includegraphics[width=.95\textwidth]{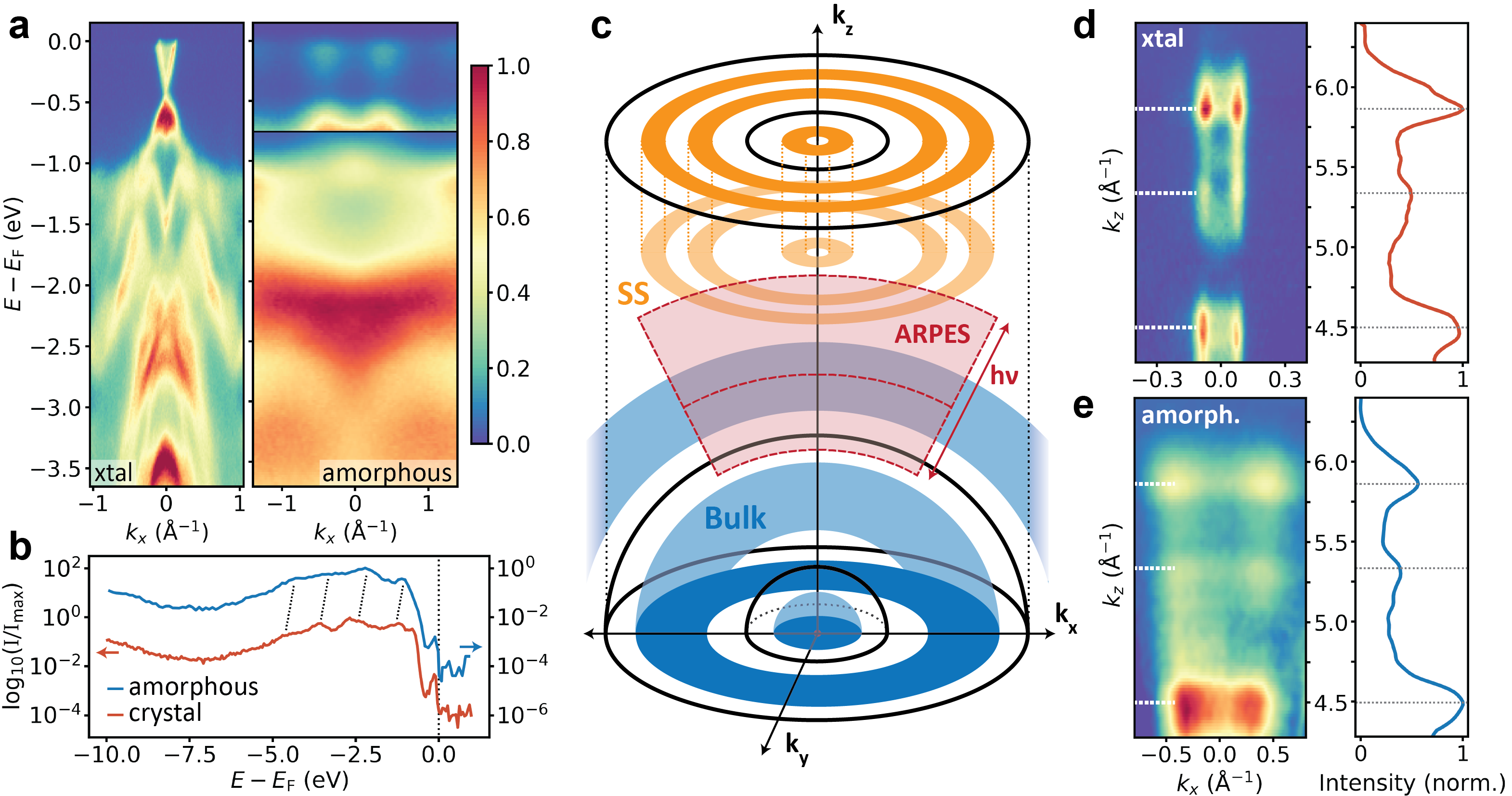}
    \caption{\textbf{Bulk and surface state comparisons to crystalline \bise{}} \textbf{(a)} Deep binding energy ARPES spectra for c-\bise{} ($\Upgamma - \mathrm{K}$) and a-\bise{}. \textbf{(b)} XPS on a-\bise{} (blue) and c-\bise{} (red) displays similar spectra for valence bands. Dashed lines indicate corresponding peaks, and the spectral hump of the upper energy portion of the surface state can be seen near \ef{} in both samples. \textbf{(c)} Diagram of amorphous band structure geometry. Bulk states form spherical shells around $\vec{k} = (0,0,0)$ (blue), whereas surface states form cylindrical shells around the $k_z$ axis (orange). For a single photon energy, ARPES probes a section of an approximately spherical shell about $\vec{k} = (0,0,0)$ (red). \textbf{(d,e)} Surface state spectrum at \ef{} as a function of $k_z$ for c-\bise{} (d) and a-\bise{} (e). $k_x$-integrated intensity shown to the left with 3 characteristic peaks marked by arrows.}
    \label{fig:fig3}
\end{figure*}

It is illustrative to compare the momentum space structure of a-\bise{} with c-\bise{} and to separate the roles of bulk and surface states.
In Fig.~\ref{fig:fig3}(a), we show deep binding-energy ARPES spectra at $h\nu = 120$ eV for c-\bise{} along the $\Upgamma-\mathrm{K}$ direction and for a-\bise{} radial from $\Upgamma$. The intensity of the features near the Fermi level in a-\bise{} has been enhanced by $10\times$ for visibility. The most notable difference close to the Fermi level is that the Dirac state in the crystal and the vertical features in the amorphous system have markedly different Fermi wave vectors (\kf{}), $0.08$ Å$^{-1}$ and $0.4$ Å$^{-1}$, respectively. We identify several factors that can contribute to this difference.
First, a surface potential (captured by $\alpha$ in Eq.~\eqref{eq:onsite}) can shift the Dirac point downwards in energy, changing $k_\mathrm{F}$ significantly. Second, the surface state Fermi velocity is not universal, and can be strongly affected by disorder, as we will exemplify later on.  

Turning back to Fig.~\ref{fig:fig3}(a), and looking deeper in binding energy we see that the crystalline sample maintains strongly dispersive features, whereas the amorphous sample demonstrates flattened and broadened bulk band structure. Notably there is a broad nearly-flat structure near $-2$ eV. 
Even though the curvature of the deep binding energy band structure is reduced in the a-\bise{}, the angle integrated spectral response is intriguingly similar. In Fig.~\ref{fig:fig3}(b) we plot the x-ray photoemission spectroscopy (XPS) spectra for c-\bise (red) and a-\bise (blue) on shifted y-axes for visibility. The overall intensity as a function of binding energy follows nearly identical large-scale behavior with a dip near $-8$ eV and a broad shoulder with substructure at $-3$ eV. The crystalline sample exhibits four peaked features in the shouldered region that correspond to four peaks in the amorphous spectrum shifted by $\sim 0.5$ eV. The similarity between the spectra indicates that the deep binding energy band structure of the amorphous sample appears as the momentum averaged band structure in the crystal.

The peculiar difference between the highly dispersive bands near the Fermi level and the weakly dispersive bands at high binding energy in a-\bise{} can be explained by reflecting on the uncommon features that a rotationally symmetric system imprints in ARPES. Fig.~\ref{fig:fig3}(c) illustrates the 3D momentum space structure of bulk bands and surface states given the symmetry of an amorphous system. The bulk bands (blue) are rotationally symmetric and are allowed to vary along the radial direction contingent on a well-defined characteristic momentum, thereby forming repeated spherical shells at constant energy within BLZs. Surface states (orange), spatially localized at the surface, are $k_z$-independent and forming cylindrical shells oriented along the $k_z$-axis. ARPES is well suited for studying $k_z$-dependence in crystals with cartesian Brillouin zones by varying the photon energy, since:
\begin{equation}
    k_z \propto \frac{1}{\hbar}\sqrt{2*m_e(h\nu - E - \phi + V_0)}.
\end{equation}
where $E$ is the binding energy of the electron, $V_0$ is the fixed inner potential of the material, $\phi$ is the material work function, $h\nu$ is the photon energy, and $m_e$ is the mass of the electron. However, in a spherically symmetric momentum-space, ARPES faces an additional challenge to observe $k_z$-dependent bulk states. In ARPES, $k_{x,y} \propto \sin{\theta}$ and $k_z \propto \cos{\theta}$, where $\theta$ is the angle of photoemission, such that ARPES probes an approximately spherical crossection in momentum space for a given photon energy. The solid red arcs in Fig.~\ref{fig:fig3}(c) illustrate variable ARPES crossections given fixed photon energies. Moreover, typical inner potentials, $V_0$, are of order 10 eV, limiting the minimum probable $k_z$ to approximately 2 Å$^{-1}$ so that the concentric structure at $k_z=0$ is out of reach. All in all, this means that ARPES probes approximately spherical cross-sections of the spherically-invariant bulk band structure. Therefore bulk dispersive features will still disperse along $k_z$-axis by varying the photon energy (shaded red region) but features in $k_x$ and $k_y$ will appear flat. 

In contrast, surface states are $k_z$ independent and the concentric structure can be accessed by ARPES at any photon energy (see red shaded region meets orange), revealing dispersive bands in $k_x$ and $k_y$. In panels (d) and (e), we display the $k_z$-dependence of the c-\bise{} and a-\bise{} surface states near \ef{}, respectively. In the crystalline case, \kf{} remains fixed for all $k_z$ at $0.08$ Å$^{-1}$, manifesting as narrow vertical pillars in $k_z$ vs. $k_x$ with variable intensity due to photon energy dependent matrix elements.  

Curiously, the amorphous surface state bands expand outward with increasing $k_z$ (Fig.~\ref{fig:fig3}(e)). This is a non-periodic dispersive feature that occurs over the full measured 2.5 inverse angstroms in $k_z$. The lack of periodicity over this range indicates that this momentum space feature cannot be due to a crystalline bulk structure since any such structure would need to repeat on smaller intervals than 2 Å ($\pi/2.25$ Å$^{-1}$). Moreover, this peculiarity does not affect the conclusion that these are surface states since, given the lack of long-range order, bulk states would necessarily form rotationally symmetric shells centered at ($k_x$, $k_y$, $k_z$) = (0, 0, 0) and would appear nearly horizontal in panel (e). Therefore these states must be of surface state origin since they are coherent across many $k_z$ values.

Crucially, the momentum integrated intensity as a function of $k_z$ (panels (e) and (d), right plots) are nearly identical, with intensity peaks occurring at the same three $k_z$ values (horizontal lines). This indicates that the photon energy dependent matrix elements are comparable in the two systems, advocating for similar orbital character.

\begin{figure*}
    \centering
    \includegraphics[width = .9 \textwidth]{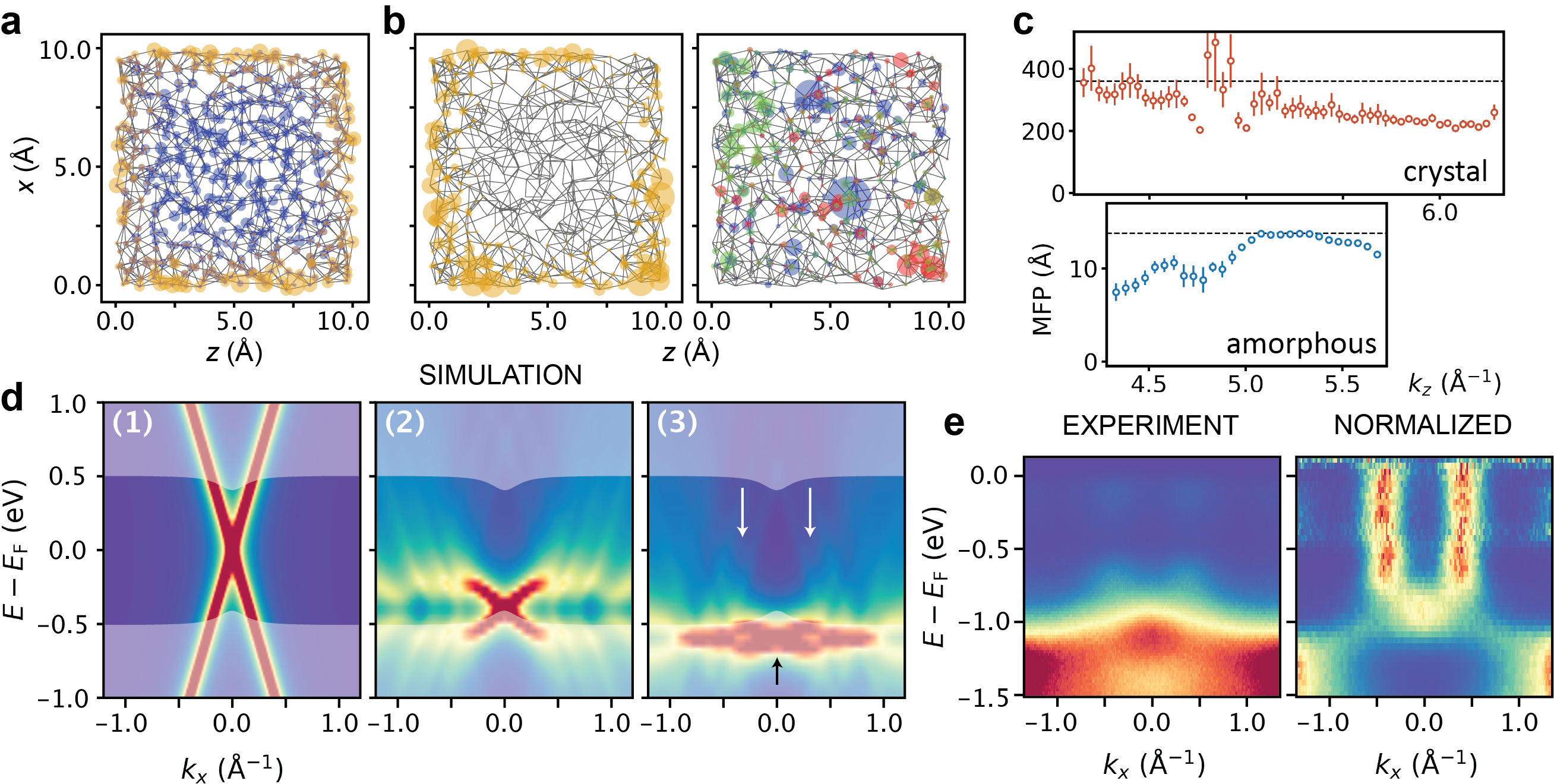}
    \caption{\textbf{(a)} Average site occupations for wavefunctions with energy within the bulk gap (yellow) and below the bulk gap (blue). \textbf{(b)} Single surface state wavefunction ($E-E_\mathrm{F} = 0$ yellow) is delocalized along the surface plane, spreading across multiple unit cells. Single bulk state wavefunctions ($E-E_\mathrm{F} < -0.5$ eV red, green blue) are localized within the bulk. \textbf{(c)} Measure of mean free-path (or spatial coherence) of surface state electrons as a function of $k_z$ as determined from $2\pi/\sigma_{\mathrm{MDC}}$ for c-\bise{} and a-\bise{} from Fig. 3(d,e). \textbf{(d)} Spectral function of a linear dispersing state scattered on a disordered array of atoms with interacting strengths (1) $v_0 = 0$, (2) $v_0 = 2$, (3) $v_0 = 3$. When the interaction with scattering centers increases, copies of the central Dirac cone appear at the peaks of $c_2$, i.e. around $k = \pm \, 0.4$ Å$^{-1}$. For strong enough scattering potential, the dispersion is pushed into the valence and vertical features form that cross the bulk band gap. \textbf{(e)} Experimental ARPES spectrum and the same spectrum normalized along the energy-axis for comparison with simulation.}
    \label{fig:fig4}
\end{figure*}

Using the BHZ model defined by Eqs.~\eqref{eq:BHZ1} and \eqref{eq:onsite}, we can compare the localization of the bulk and surface wavefunctions. Fig.~\ref{fig:fig4}(a) shows the average wavefunction site occupation within a 2 Å slice of the amorphous cube for surface states between $E-E_\mathrm{D} = -1.0$ and 0 eV (yellow) and bulk states between $E-E_\mathrm{D} = -9.5$ and $-4.0$ eV (blue), where $E_\mathrm{D}$ marks the center of the band gap. The in-gap surface states are localized to the system edges whereas the bulk states evenly fill the interior appearing completely delocalized. However, the singular wavefunctions tell a different story. Fig.~\ref{fig:fig4}(b) shows a single wavefunction at $-0.45$ eV (yellow), again localized to the surface but delocalized along the 2D surfaces. A random selection of three bulk wavefunctions between $-9.5$ and $4.0$ (red, green, and blue) show a localized behavior, constrained to a small number of sites.

In fact, we are able to deduce a lower limit on the coherence length of the electronic order from the momentum broadening of the states in Fig.~\ref{fig:fig3}(d) \cite{Robarts2020,Baumberger2004,Kim2011}. The momentum Lorentzian linewidth of the two peaks in each case serve as a measure of the electronic real space ordering, in so much as perfect electronic order leads to delta functions in momentum (ignoring lifetime broadening) and spatial incoherence leads to smearing across the full BZ. This represents a lower limit for the coherence length since final state effects, spatial variations in doping, or lifetime broadening could introduce additional extrinsic broadening. 

Fig.~\ref{fig:fig4}(c) plots the $k_z$-dependence of the coherence length (or mean free path, MFP) as $2\pi/\Gamma_k$ where $\Gamma_k$ is the Lorentzian line-width. For c-\bise{}, MFP is largest at small $k_z$ at nearly 400 Å, serving as our estimate of the lower limit of the coherence length in the crystal. The MFP then linearly decreases towards larger $k_z$ suggesting a possible photon energy dependence of the momentum resolution. In a-\bise{} the largest calculated MFP is near $k_z\approx$ 5.2 Å$^{-1}$ which gives a value for the MFP of 13 Å. This is a markedly reduced coherence to the crystal yet coherent beyond three nearest neighbors, $3a^*$. Interestingly, the measured coherence length reduces towards lower and higher $k_z$ values indicating an additional broadening mechanism in the low momentum regime as compared to c-\bise{}.

The delocalization of the surface state along the surface planes enables the electronic wavefunction to encompass many atomic sites, which in turn enables dispersive coherent structure in momentum space. In contrast, the localization of the bulk bands, inferred from Fig.~\ref{fig:fig4}(b) would suggest a flattened band structure. This is indeed the case for the amorphous structure in Fig.~\ref{fig:fig3}(a), in which strong dispersion occurs for the surface state near $E_\mathrm{F}$ and bulk bands deeper in binding energy are flat. It is possible that either topological protection or spin-momentum locking of the surface states may enhance the in-plane coherence for a-\bise{} surface states, making it a particular good material for observing a defined amorphous band structure.

Lastly we discuss why the amorphous surface state may exhibit broad near-vertical dispersive features at a dramatically enhanced $k_\mathrm{F}$ with respect to the crystalline spectrum (see Fig.~\ref{fig:fig3}(a)). These broad vertical features, particular to a-\bise{}, are not fully explained by our tight-binding model used in Fig.~\ref{fig:fig2}. A tantalizing additional effect neglected in the tight-binding approximation is that scattering on a disordered array of atoms can significantly alter the dispersion of a propagating state, creating the effect of a broad and vertical dispersion~\cite{Edwards1961, Edwards1962}. This is believed to be the case of liquid metals and surface electrons propagating within a disordered but correlated array of atoms on crystal surfaces~\cite{Ryu2021}. Translated to our situation, the hypothesis is that the surface Dirac propagating state experiences scattering due to the correlated disorder intrinsic to amorphous structure. This weak scattering effect is not captured by the tight-binding approximation~\cite{Edwards1961, Edwards1962}. To explore this possibility, we calculate the disorder induced self-energy caused by the experimental radial density function of Fig.~\ref{fig:fig1}(d) to determine how a linearly dispersing surface state is affected by this spatial distribution of atoms. 
Following Ref.~\cite{Edwards1961}, the scattering effect introduces a self-energy that reads
\begin{equation}
    \Sigma_s(\mathbf{k}) = v_0\sum_{s'}\int c_2(\mathbf{k}-\mathbf{k}')F_{ss'}(\mathbf{k},\mathbf{k}')G_{0s'}(\mathbf{k}')\mathrm{d}\mathbf{k}', \label{eq:Edwards}
\end{equation}
where $v_0$ is the strength of the disorder, $F_{ss'}(\mathbf{k},\mathbf{k'})$ is the overlap factor between the two bands (labeled by $s=\pm$) of the Dirac Hamiltonian, $c_2(\mathbf{k})$ is the Fourier transform of the radial distribution function shown in Fig.~\ref{fig:fig1}(d), and $G_{0s} = (E-sv_F|\mathbf{k}|+i0^{+})^{-1}$ is the bare Green's Function of the surface Dirac cone in the diagonal basis.

Fig.~\ref{fig:fig4}(d) shows that indeed correlated structural disorder reshapes the linear dispersion for increasing disorder strengths $v_0$. Without interaction, the spectral function only shows a Dirac cone centered around $\Gamma$. When the interaction with scattering centers increases, copies of the central Dirac cone appear at the peaks of $c_2$, i.e. around $k = \pm 0.4$ Å$^{-1}$. For strong enough scattering potential, the dispersion and node (black arrow) is pushed into the valence (shaded white region) and vertical features form that cross the bulk band gap (white arrows). Thus the exact shape of the dispersion can be strongly affected by the surface disorder, a mechanism that could explain why the ARPES spectrum measured in Fig.~\ref{fig:fig3} differs significantly from the crystalline spectrum.

Comparing Fig.~\ref{fig:fig4}(d) with Fig.~\ref{fig:fig4}(e) (where the ARPES spectrum near $E_\mathrm{F}$ is shown along with the spectrum normalized along binding energy) demonstrates how both the experiment and simulation exhibit broad vertical features that cross the gap, as well as an apparent band crossing and node at the valence band edge. 

\section*{Discussion}

Our data indicates several directions that merit future study. First, the features we see arise from well defined short-range length scales, hinting at the possibility of widely overlooked momentum-space structure in all non-crystalline solids with this property, such as other amorphous materials, quasicrystals, and liquids. Diagnosing them with further photoemission studies complemented by other probes, such as scanning tunneling spectroscopy, has the potential to change significantly the landscape of solid-state properties. The presence of highly dispersive features in a-\bise{} motivate the generalization of momentum-dependent phenomena to glassy systems, including spin-momentum locking~\cite{Corbae2023}, momentum-based paring in superconductivity, or new avenues to engineer flat-bands in amorphous phase-change materials~\cite{Marsal2022}.

Based on a-\bise{} specifically, another direction of further study is the origin of the appreciable monotonic $k_z$-dependence of the surface states of Fig.~\ref{fig:fig3}(d), which clearly differs from bulk states, but is not completely $k_z$-independent.
A possible cause is the lack of translational symmetry itself. ARPES is based on the notion that translational symmetry conserves the in-plane crystalline momentum following photoemission. This may no longer be the case in amorphous samples; while continuous translations can be recovered on average to explain most of our results, discrete translational symmetry is lost, which may result in more subtle $k_z$ dependencies, and may be a unique feature of non-crystalline media.

\section*{Conclusion}

In summary, we have revealed highly dispersive surface electronic states on the surface of amorphous \bise{}, that exhibit a rotationally symmetric Fermi surface with repeated Brillouin zone-like repetitions. This is made possible by the presence of a well-defined real-space length scale from the disordered hard packing of atoms in the amorphous structure, which corresponds to a well-defined reciprocal length scale. The amorphous analog to crystalline \bise{} preserves the angle-integrated XPS spectral features yet exhibits remarkably different valence state behavior, manifesting as vertical-like features with large Fermi wave-vectors.  
Since the presence of local chemical order is ubiquitous in solids, our work calls for a retrospective investigation of amorphous and other non-crystalline systems such as quasicrystals, in search for dispersive features that reveal novel quantum effects in momentum space, previously reserved for crystals alone.

\section*{Methods}

ARPES spectra were acquired from the MAESTRO $\upmu$ARPES endstation (BL 7.0.2.1) and the MERLIN ARPES endstation (BL 4.0.3) at the Advanced Light Source at Lawrence Berkeley National Laboratory with photon energies between 60 and 140 eV and at temperatures below 80 K. Samples for ARPES were capped with 50 nm of selenium immediately following growth and decapped in 10$^{-11}$ Torr base pressure 
directly before measurement. We observed Fresnel color changes of the surface and waited for the color changing stopped to ensure there is no residual selenium.
We confirmed that there is not residual selenium through angle dependent XPS. Furthermore, the decap procedure does not cause crystallization of the films \cite{Corbae2023}.

\section*{Acknowledgements}
We are grateful to S. Ciuchi, S. Franca, S. Fratini, D. Mayou, D. Mu\~{n}oz-Segovia and S. Tchoumakov for discussions.
This work was primarily supported by the Director, Office of Science, Office of Basic Energy Sciences, Materials Sciences and Engineering Division, of the U.S. Department of Energy, under Contract No. DE-AC02-05CH11231, as part of the Ultrafast Materials Science Program (KC2203) with secondary contributions from the Nonequilibrium Magnetism Program (KC2204) and the Electronic Materials Program (EMAT). TEM work at the Molecular Foundry was supported by the Office of Science, Office of Basic Energy Sciences, of the U.S. Department of Energy under Contract No. DE-AC02-05CH11231.
A.G.G. and Q. M acknowledge financial support from the European Union Horizon 2020 research and innovation program under grant agreement No. 829044 (SCHINES). A.G.G. is also supported by the European Research Council (ERC) Consolidator grant under grant agreement No. 101042707 (TOPOMORPH).

\section*{Author Contributions}
S.C. performed the ARPES measurements and the analysis. 
A.L. and S.C. developed the experimental premise and infrastructure.
Q.M. and A.G.G. devised the theoretical modelling. Q. M. carried out the numerical and analytical calculations supervised by D. V. and A. G. G. 
The initial idea for this experiment came from PC and FH.  PC grew and characterized the films (XPS, Raman) supervised by and in consultation with F.H, and participated in ARPES measurements of the films. EK performed and analyzed the HRTEM, supervised by and in consultation with MS.
All authors contributed to writing the manuscript.

\bibliographystyle{naturemag}
\bibliography{abib.bib}

\begin{thebibliography}{10}
\expandafter\ifx\csname url\endcsname\relax
  \def\url#1{\texttt{#1}}\fi
\expandafter\ifx\csname urlprefix\endcsname\relax\def\urlprefix{URL }\fi
\providecommand{\bibinfo}[2]{#2}
\providecommand{\eprint}[2][]{\url{#2}}

\bibitem{Ashcroft76}
\bibinfo{author}{Ashcroft, N.~W.} \& \bibinfo{author}{Mermin, N.~D.}
\newblock \emph{\bibinfo{title}{{S}olid {S}tate {P}hysics}}
  (\bibinfo{publisher}{Holt-Saunders}, \bibinfo{year}{1976}).

\bibitem{Anderson1958}
\bibinfo{author}{Anderson, P.~W.}
\newblock \bibinfo{title}{Absence of diffusion in certain random lattices}.
\newblock \emph{\bibinfo{journal}{Phys. Rev.}} \textbf{\bibinfo{volume}{109}},
  \bibinfo{pages}{1492--1505} (\bibinfo{year}{1958}).

\bibitem{Mott1978}
\bibinfo{author}{Mott, S.~N.}
\newblock \bibinfo{title}{Electrons in glass}.
\newblock \emph{\bibinfo{journal}{Rev. Mod. Phys.}}
  \textbf{\bibinfo{volume}{50}}, \bibinfo{pages}{203--208}
  (\bibinfo{year}{1978}).

\bibitem{Ludlam2005}
\bibinfo{author}{Ludlam, J.~J.}, \bibinfo{author}{Taraskin, S.~N.},
  \bibinfo{author}{Elliott, S.~R.} \& \bibinfo{author}{Drabold, D.~A.}
\newblock \bibinfo{title}{Universal features of localized eigenstates in
  disordered systems}.
\newblock \emph{\bibinfo{journal}{Journal of Physics: Condensed Matter}}
  \textbf{\bibinfo{volume}{17}}, \bibinfo{pages}{L321--L327}
  (\bibinfo{year}{2005}).

\bibitem{Mitchell2018}
\bibinfo{author}{Mitchell, N.~P.}, \bibinfo{author}{Nash, L.~M.},
  \bibinfo{author}{Hexner, D.}, \bibinfo{author}{Turner, A.~M.} \&
  \bibinfo{author}{Irvine, W. T.~M.}
\newblock \bibinfo{title}{Amorphous topological insulators constructed from
  random point sets}.
\newblock \emph{\bibinfo{journal}{Nature Physics}}
  \textbf{\bibinfo{volume}{14}} (\bibinfo{year}{2018}).

\bibitem{Varjas2019}
\bibinfo{author}{Varjas, D.} \emph{et~al.}
\newblock \bibinfo{title}{Topological phases without crystalline counterparts}.
\newblock \emph{\bibinfo{journal}{Phys. Rev. Lett.}}
  \textbf{\bibinfo{volume}{123}}, \bibinfo{pages}{196401}
  (\bibinfo{year}{2019}).

\bibitem{Robarts2020}
\bibinfo{author}{Robarts, H.~C.} \emph{et~al.}
\newblock \bibinfo{title}{Extreme {F}ermi surface smearing in a maximally
  disordered concentrated solid solution}.
\newblock \emph{\bibinfo{journal}{Phys. Rev. Lett.}}
  \textbf{\bibinfo{volume}{124}}, \bibinfo{pages}{046402}
  (\bibinfo{year}{2020}).

\bibitem{Ding2015}
\bibinfo{author}{Ding, J.}, \bibinfo{author}{Ma, E.}, \bibinfo{author}{Asta,
  M.} \& \bibinfo{author}{Ritchie, R.~O.}
\newblock \bibinfo{title}{Second-nearest-neighbor correlations from connection
  of atomic packing motifs in metallic glasses and liquids}.
\newblock \emph{\bibinfo{journal}{Scientific Reports}}
  \textbf{\bibinfo{volume}{5}}, \bibinfo{pages}{17429} (\bibinfo{year}{2015}).

\bibitem{zallen}
\bibinfo{author}{Zallen, R.}
\newblock \emph{\bibinfo{title}{The Physics of Amorphous Solids}}
  (\bibinfo{publisher}{Wiley}, \bibinfo{year}{1998}).

\bibitem{Zachariasen1932}
\bibinfo{author}{Zachariasen, W.~H.}
\newblock \bibinfo{title}{The atomic arrangement in glass}.
\newblock \emph{\bibinfo{journal}{Journal of the American Chemical Society}}
  \textbf{\bibinfo{volume}{54}}, \bibinfo{pages}{3841--3851}
  (\bibinfo{year}{1932}).

\bibitem{Vink2001}
\bibinfo{author}{Vink, R. L.~C.}, \bibinfo{author}{Barkema, G.~T.},
  \bibinfo{author}{Stijnman, M.~A.} \& \bibinfo{author}{Bisseling, R.~H.}
\newblock \bibinfo{title}{Device-size atomistic models of amorphous silicon}.
\newblock \emph{\bibinfo{journal}{Phys. Rev. B}} \textbf{\bibinfo{volume}{64}},
  \bibinfo{pages}{245214} (\bibinfo{year}{2001}).

\bibitem{Deringer2018}
\bibinfo{author}{Deringer, V.~L.} \emph{et~al.}
\newblock \bibinfo{title}{Realistic atomistic structure of amorphous silicon
  from machine-learning-driven molecular dynamics}.
\newblock \emph{\bibinfo{journal}{The Journal of Physical Chemistry Letters}}
  \textbf{\bibinfo{volume}{9}}, \bibinfo{pages}{2879--2885}
  (\bibinfo{year}{2018}).

\bibitem{Corbae2023}
\bibinfo{author}{Corbae, P.} \emph{et~al.}
\newblock \bibinfo{title}{Observation of spin-momentum locked surface states in
  amorphous {B}i$_2${S}e$_3$}.
\newblock \emph{\bibinfo{journal}{Nature Materials}}  (\bibinfo{year}{2023}).

\bibitem{Ryu2021}
\bibinfo{author}{Ryu, S.~H.} \emph{et~al.}
\newblock \bibinfo{title}{Pseudogap in a crystalline insulator doped by
  disordered metals}.
\newblock \emph{\bibinfo{journal}{Nature}} \textbf{\bibinfo{volume}{596}},
  \bibinfo{pages}{68--73} (\bibinfo{year}{2021}).

\bibitem{Ziman1961}
\bibinfo{author}{Ziman, J.~M.}
\newblock \bibinfo{title}{A theory of the electrical properties of liquid
  metals. {I}: {T}he monovalent metals}.
\newblock \emph{\bibinfo{journal}{The Philosophical Magazine: A Journal of
  Theoretical Experimental and Applied Physics}} \textbf{\bibinfo{volume}{6}},
  \bibinfo{pages}{1013--1034} (\bibinfo{year}{1961}).

\bibitem{Edwards1961}
\bibinfo{author}{Edwards, S.~F.}
\newblock \bibinfo{title}{The electronic structure of disordered systems}.
\newblock \emph{\bibinfo{journal}{The Philosophical Magazine: A Journal of
  Theoretical Experimental and Applied Physics}} \textbf{\bibinfo{volume}{6}},
  \bibinfo{pages}{617--638} (\bibinfo{year}{1961}).

\bibitem{Edwards1962}
\bibinfo{author}{Edwards, S.~F.} \& \bibinfo{author}{Mott, N.~F.}
\newblock \bibinfo{title}{The electronic structure of liquid metals}.
\newblock \emph{\bibinfo{journal}{Proceedings of the Royal Society of London.
  Series A. Mathematical and Physical Sciences}}
  \textbf{\bibinfo{volume}{267}}, \bibinfo{pages}{518--540}
  (\bibinfo{year}{1962}).

\bibitem{Morgan1969}
\bibinfo{author}{Morgan, G.~J.}
\newblock \bibinfo{title}{Electron transport in liquid metals {I}{I}. {A} model
  for the wave functions in liquid transition metals}.
\newblock \emph{\bibinfo{journal}{Journal of Physics C: Solid State Physics}}
  \textbf{\bibinfo{volume}{2}}, \bibinfo{pages}{1454--1464}
  (\bibinfo{year}{1969}).

\bibitem{Gyorffy1970}
\bibinfo{author}{Gyorffy, B.~L.}
\newblock \bibinfo{title}{Electronic states in liquid metals: A generalization
  of the coherent-potential approximation for a system with short-range order}.
\newblock \emph{\bibinfo{journal}{Phys. Rev. B}} \textbf{\bibinfo{volume}{1}},
  \bibinfo{pages}{3290--3299} (\bibinfo{year}{1970}).

\bibitem{Schwartz1971}
\bibinfo{author}{Schwartz, L.} \& \bibinfo{author}{Ehrenreich, H.}
\newblock \bibinfo{title}{Single-site approximations in the electronic theory
  of liquid metals}.
\newblock \emph{\bibinfo{journal}{Annals of Physics}}
  \textbf{\bibinfo{volume}{64}}, \bibinfo{pages}{100--148}
  (\bibinfo{year}{1971}).

\bibitem{Chang1975}
\bibinfo{author}{Chang, K.~S.}, \bibinfo{author}{Sher, A.},
  \bibinfo{author}{Petzinger, K.~G.} \& \bibinfo{author}{Weisz, G.}
\newblock \bibinfo{title}{Density of states of liquid cu}.
\newblock \emph{\bibinfo{journal}{Phys. Rev. B}} \textbf{\bibinfo{volume}{12}},
  \bibinfo{pages}{5506--5513} (\bibinfo{year}{1975}).

\bibitem{Olson1975}
\bibinfo{author}{Olson, J.~J.}
\newblock \bibinfo{title}{Anderson-mcmillan prescription for the density of
  states of liquid iron}.
\newblock \emph{\bibinfo{journal}{Phys. Rev. B}} \textbf{\bibinfo{volume}{12}},
  \bibinfo{pages}{2908--2916} (\bibinfo{year}{1975}).

\bibitem{Oglesby1976}
\bibinfo{author}{Oglesby, J.} \& \bibinfo{author}{Lloyd, P.}
\newblock \bibinfo{title}{Some single-site structure-independent approximations
  in condensed materials}.
\newblock \emph{\bibinfo{journal}{Journal of Physics C: Solid State Physics}}
  \textbf{\bibinfo{volume}{9}}, \bibinfo{pages}{2879--2886}
  (\bibinfo{year}{1976}).

\bibitem{Ridgway1998}
\bibinfo{author}{Ridgway, M.~C.}, \bibinfo{author}{Glover, C.~J.},
  \bibinfo{author}{Foran, G.~J.} \& \bibinfo{author}{Yu, K.~M.}
\newblock \bibinfo{title}{Characterization of the local structure of amorphous
  {G}a{A}s produced by ion implantation}.
\newblock \emph{\bibinfo{journal}{Journal of Applied Physics}}
  \textbf{\bibinfo{volume}{83}}, \bibinfo{pages}{4610--4614}
  (\bibinfo{year}{1998}).

\bibitem{Onodera2020}
\bibinfo{author}{Onodera, Y.} \emph{et~al.}
\newblock \bibinfo{title}{Structure and properties of densified silica glass:
  characterizing the order within disorder}.
\newblock \emph{\bibinfo{journal}{NPG Asia Materials}}
  \textbf{\bibinfo{volume}{12}}, \bibinfo{pages}{85} (\bibinfo{year}{2020}).

\bibitem{toh2020}
\bibinfo{author}{Toh, C.-T.} \emph{et~al.}
\newblock \bibinfo{title}{Synthesis and properties of free-standing monolayer
  amorphous carbon}.
\newblock \emph{\bibinfo{journal}{Nature}} \textbf{\bibinfo{volume}{577}},
  \bibinfo{pages}{199--203} (\bibinfo{year}{2020}).

\bibitem{Perret2021}
\bibinfo{author}{Perret, E.} \& \bibinfo{author}{Hufenus, R.}
\newblock \bibinfo{title}{Insights into strain-induced solid mesophases in
  melt-spun polymer fibers}.
\newblock \emph{\bibinfo{journal}{Polymer}} \textbf{\bibinfo{volume}{229}},
  \bibinfo{pages}{124010} (\bibinfo{year}{2021}).

\bibitem{Zhang2009}
\bibinfo{author}{Zhang, H.} \emph{et~al.}
\newblock \bibinfo{title}{Topological insulators in {B}i$_2${S}e$_3$,
  {B}i$_2${T}e$_3$ and {S}b$_2${T}e$_3$ with a single {D}irac cone on the
  surface}.
\newblock \emph{\bibinfo{journal}{Nature Physics}}
  \textbf{\bibinfo{volume}{5}}, \bibinfo{pages}{438--442}
  (\bibinfo{year}{2009}).

\bibitem{Baumberger2004}
\bibinfo{author}{F., B.}, \bibinfo{author}{W., A.}, \bibinfo{author}{T., G.} \&
  \bibinfo{author}{J., O.}
\newblock \bibinfo{title}{Electron coherence in a melting lead monolayer}.
\newblock \emph{\bibinfo{journal}{Science}} \textbf{\bibinfo{volume}{306}},
  \bibinfo{pages}{2221--2224} (\bibinfo{year}{2004}).

\bibitem{Kim2011}
\bibinfo{author}{Kim, K.~S.} \& \bibinfo{author}{Yeom, H.~W.}
\newblock \bibinfo{title}{Radial band structure of electrons in liquid metals}.
\newblock \emph{\bibinfo{journal}{Phys. Rev. Lett.}}
  \textbf{\bibinfo{volume}{107}}, \bibinfo{pages}{136402}
  (\bibinfo{year}{2011}).

\bibitem{Marsal2022}
\bibinfo{author}{{Marsal}, Q.}, \bibinfo{author}{{Varjas}, D.} \&
  \bibinfo{author}{{Grushin}, A.~G.}
\newblock \bibinfo{title}{{Obstructed insulators and flat bands in topological
  phase-change materials}}.
\newblock \emph{\bibinfo{journal}{arXiv e-prints}}
  \bibinfo{pages}{arXiv:2204.14177} (\bibinfo{year}{2022}).
\newblock \eprint{2204.14177}.

\end{thebibliography}

\end{document}